\documentstyle[emulateapj]{article}

\submitted{to appear in the Astrophysical Journal, Part 1}

\lefthead{Hachisu and Kato}
\righthead{Recurrent Nova: V394 Coronae Austrinae}

\begin{document}

\title{A THEORETICAL LIGHT-CURVE MODEL FOR THE RECURRENT NOVA 
V394 CORONAE AUSTRINAE}

\author{Izumi Hachisu}
\affil{Department of Earth Science and Astronomy, 
College of Arts and Sciences, University of Tokyo,
Komaba, Meguro-ku, Tokyo 153-8902, Japan; 
hachisu@chianti.c.u-tokyo.ac.jp}

\and

\author{Mariko Kato}
\affil{Department of Astronomy, Keio University, 
Hiyoshi, Kouhoku-ku, Yokohama 223-8521, Japan; 
mariko@educ.cc.keio.ac.jp}




\begin{abstract}
     A theoretical light curve for the 1987 outburst 
of V394 Coronae Austrinae (V394 CrA) is modeled 
to obtain various physical parameters 
of this recurrent nova.  We then apply the same set of parameters
to a quiescent phase and confirm that these parameters give 
a unified picture of the binary.
Our V394 CrA model consists of a very massive white dwarf (WD) 
with an accretion disk (ACDK) having a flaring-up rim, 
and a lobe-filling, slightly evolved, 
main-sequence star (MS).  The model includes irradiation effects 
of the MS and the ACDK by the WD.  
The early visual light curve ($t \sim 1-10$ days after the optical 
maximum) is well reproduced by a thermonuclear runaway model 
on a very massive WD close to the Chandrasekhar limit
($1.37 \pm 0.01 ~M_\odot$).
The ensuing plateau phase ($t \sim 10-30$ days) is also reproduced 
by the combination of a slightly irradiated MS 
and a fully irradiated flaring-up disk with a radius 
$\sim 1.4$ times the Roche lobe size.
The best fit parameters are the WD mass 
$\sim 1.37 ~M_\odot$, the companion mass 
$\sim 1.5 M_\odot$ ($0.8-2.0 M_\odot$ is acceptable),
the inclination angle of the orbit $i \sim $65---68$\arcdeg$, and 
the flaring-up rim $\sim 0.30$ times the disk radius.
The envelope mass at the optical peak is estimated to be 
$\sim 6 \times 10^{-6} M_\odot$, which indicates
an average mass accretion rate of
$\sim 1.5 \times 10^{-7} M_\odot$ yr$^{-1}$ 
during the quiescent phase between the 1949 and 1987 outbursts.
In the quiescent phase, we properly include 
an accretion luminosity of the WD and a viscous luminosity of the ACDK 
as well as the irradiation effects of the ACDK and MS by the WD.
The observed light curve can be reproduced with a disk size of
0.7 times the Roche lobe size and a rather slim thickness of
0.05 times the accretion disk size at the rim.  About 0.5 mag
sinusoidal variation of the light curve requires the mass accretion 
rate higher than $\sim 1.0 \times 10^{-7} M_\odot$ yr$^{-1}$,
which is consistent with the above estimation from the 1987 outburst.
These newly obtained quantities are exactly the same as 
those predicted in a new progenitor model of Type Ia supernovae. 
\end{abstract}




%

\section{INTRODUCTION}
Type Ia supernovae (SNe Ia) are one of the most luminous explosive events 
of stars.  Recently, SNe Ia have been used as good distance indicators
which provide a promising tool for determining cosmological parameters
because of their almost uniform maximum luminosities
(\cite{rie98}; \cite{per99}).
These both groups derived the maximum luminosities ($L_{\rm max}$)
of SNe Ia completely empirically from the shape of
the light curve (LCS) of nearby SNe Ia, and assumed
that the same $L_{\rm max}$--LCS relation holds for high red-shift
SNe Ia.  To be sure of any systematic biases, the physics
of SNe Ia must be understood completely.  By far, one of
the greatest problems facing SN Ia theorists is the lack of a real
progenitor (e.g., \cite{liv99} for a recent review).  
Finding a reliable progenitor is urgently required in SN Ia research.
Recurrent novae are probably the best candidate for this target
(e.g., Starrfield, Sparks, \& Truran 1985; Hachisu et al. 1999b;
Hachisu, Kato, \& Nomoto 1999a).   

Recently, the recurrent nova U Sco underwent the sixth recorded 
outburst on February 25, 1999. 
For the first time, a complete light curve has been obtained
from the rising phase to the final fading phase 
toward quiescence through the mid-plateau phase 
(e.g., Matsumoto, Kato, \& Hachisu 2000).   
Constructing a theoretical light curve of the outburst,
Hachisu et al. (2000a) have estimated various physical parameters 
of U Sco:  (1) The early linear phase of the outburst
($t \sim 1-10$ days after the optical maximum) is well reproduced 
by a thermonuclear runaway model on a $1.37 \pm 0.01 M_\odot$ white
dwarf (WD).  (2) The envelope mass at the optical maximum is estimated
to be $\sim 3 \times 10^{-6} M_\odot$, which results in the mass 
transfer rate of $\sim 2.5 \times 10^{-7} M_\odot$ yr$^{-1}$
during the quiescent phase between the 1987 and 1999 outbursts.
(3) About 60\% of the envelope mass has been blown off in the outburst
wind but the residual 40\% ($\sim 1.2 \times 10^{-6} M_\odot$)
of the envelope mass has been left and accumulated on the white dwarf.
Therefore, the net mass increasing rate of the white dwarf
is $\sim 1.0 \times 10^{-7} M_\odot$ yr$^{-1}$, which meets
the condition for SN Ia explosions 
of carbon-oxygen cores (\cite{nom91}).
Thus, Hachisu et al. (2000a, 2000b) have concluded that 
the white dwarf mass in U Sco will reach the critical
mass ($M_{\rm Ia}= 1.378 M_\odot$,  
taken from Nomoto, Thielemann, \& Yokoi 1984)
in quite a near future and explode as an SN Ia.  
Therefore, we regard that U Sco is 
a very strong candidate for the immediate progenitor of SNe Ia.

     It has been suggested that the recurrent nova V394 CrA 
is a twin system of U Sco because of its almost same decline rate 
of the early light curve and spectrum feature during the outburst
(e.g., \cite{sek89}).  It is very likely that the physical 
parameters obtained for U Sco are common to V394 CrA.  In this
paper, we derive various physical quantities of V394 CrA,
both during the 1987 outburst and in quiescence, by constructing 
the same theoretical light curve models as for U Sco, and 
examine whether or not V394 CrA is an immediate progenitor of 
SNe Ia.  In \S 2, we briefly describe our light curve
model during the outburst and present the fitting results for the 1987
outburst of V394 CrA.  In \S 3, based on the physical parameters 
obtained in \S 2, we construct theoretical light curves 
for the quiescent phase of V394 CrA and confirm that the parameters 
during the outburst are consistent with those in quiescence.
Discussion follows in \S 4, especially for relevance to SN Ia
progenitors.

\section{LIGHT CURVES FOR THE 1987 OUTBURST}
\begin{figure}
\plotone{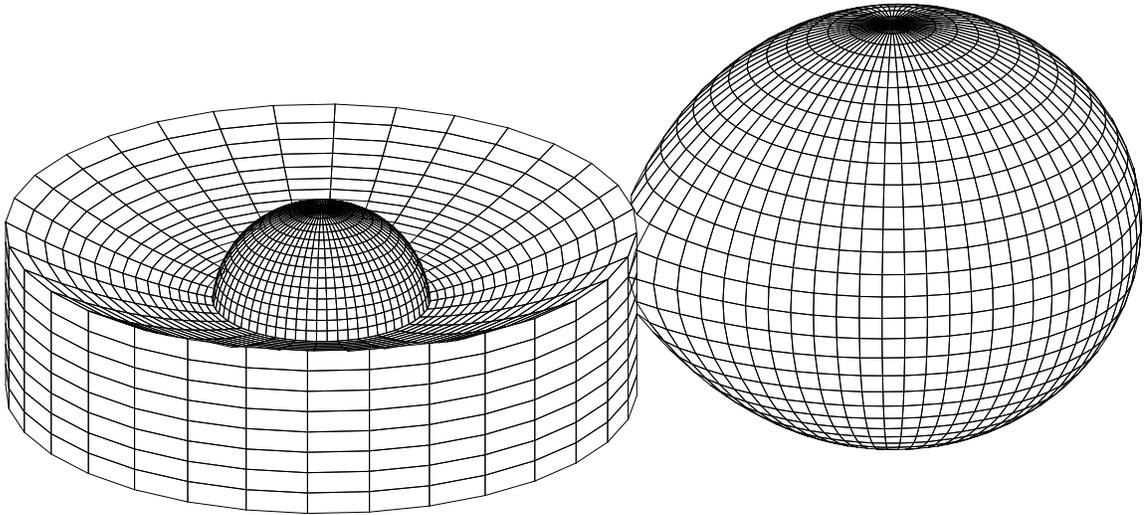}
\caption{
Configuration of our V394 CrA model in the 1987 outburst 
$\sim 11$ days after the optical maximum, i.e., 
when the WD photosphere shrinks to $R_{\rm ph} = 1.0 R_\odot$.
The cool component (right figure) is a slightly evolved MS
($1.5 M_\odot$) filling up its inner critical Roche lobe.  
Only the north and south polar areas of the secondary are 
heated up by the hot component ($1.37 ~M_\odot$ WD, left figure) 
because a large part of the light from the hot component 
is blocked by the flaring-up edge of the accretion disk.  
Here, the separation is $a= 4.97 R_\odot$, and
the effective radii of the inner critical Roche lobes are
$R_1^*= 1.84 R_\odot$, and $R_2^*= R_2= 1.92 R_\odot$, 
for the primary WD and the secondary MS, respectively.
The inclination angle is $i = 68\arcdeg$.  The disk size and
thickness are $\alpha=1.4$ and $\beta=0.30$, respectively.
\label{v394cra87_fig_burst}}
\end{figure}

     The orbital period of V394 CrA has been determined 
to be $P= 0.7577$ days by Schaefer (1990).  Here, we adopt
the ephemeris of HJD 2,447,000.250+0.7577$\times E$
at the epoch of the main-sequence companion in front.  
The orbit of the companion star is assumed to be circular. 
Our theoretical light curve model has already been described 
in Hachisu et al. (2000a) for outburst phases 
and in Hachisu et al. (2000b) for quiescent phases of U Sco. 
The total visual light is calculated from three components 
of the system: (1) the WD photosphere, (2) the MS photosphere, 
which fills its Roche lobe, and (3) the accretion disk (ACDK) 
surface, the size and thickness of which are simply defined 
by two parameters, $\alpha$ and $\beta$, as 
\begin{equation}
R_{\rm disk} = \alpha R_1^*,
\label{accretion-disk-size}
\end{equation}
and
\begin{equation}
h = \beta R_{\rm disk} \left({{\varpi} 
\over {R_{\rm disk}}} \right)^\nu,
\label{flaring-up-disk}
\end{equation}
where $R_{\rm disk}$ is the outer edge of the ACDK,
$R_1^*$ the effective radius of the inner critical Roche lobe 
for the WD component, and
$h$ the height of the surface from the equatorial plane,
$\varpi$ the distance on the equatorial plane 
from the center of the WD as seen in Figure \ref{v394cra87_fig_burst}.
Here, we adopt $\varpi$-squared law ($\nu=2$), otherwise specified,
to mimic the effect of flaring-up at the rim of the ACDK
(e.g., Schandl, Meyer-Hofmeister, \& Meyer 1997). 

\placefigure{v394cra87_fig_burst}

     It has been established that the WD photosphere expands
to a giant size at the optical maximum and then it decays
gradually to the original size of the WD in quiescence,
with the bolometric luminosity being kept near the Eddington 
luminosity (e.g., Starrfield, Sparks, \& Shaviv 1988). 
An optically thick wind is blowing from the WD during 
the outburst, which plays a key role in determining the
nova duration because a large part of the envelope mass is
carried away by the wind.
The development of the WD photosphere during the outburst
is followed by a series of optically thick wind solutions
(\cite{kat94}).
Each envelope solution is uniquely specified by the envelope mass, 
which is decreasing in time due to wind and nuclear burning.   
We have calculated solutions for five WD masses of 
$M_{\rm WD}= 1.377$, 1.37, 1.36, 1.35, and 1.3 $M_\odot$
with different hydrogen contents of $X=$0.04, 0.05, 0.06, 0.07,
0.08, 0.10, and 0.15, in mass weight.  
The numerical method and physical properties 
of the solutions are described in Kato and Hachisu (1994), 
and also in Kato (1999), where we use the revised OPAL opacity 
(\cite{igl96}).  

Assuming a blackbody photosphere of the WD envelope, we have
estimated the visual magnitude with a response function given by 
Allen (1973).   
For simplicity, we do not consider the limb-darkening effect.
The early 10 days light curve of V394 CrA is mainly determined
by the WD photosphere because the photospheric radius is larger
than or as large as the binary size and much brighter than
the MS and the ACDK.  The decline rate during the early 10 days 
($t \sim 1$---10 days after the optical maximum) 
depends very sensitively on the WD mass but hardly on the hydrogen
content (\cite{kat99}) 
or on the MS mass as described in Hachisu et al. (2000a).  
Thus, we obtain $M_{\rm WD}=1.37 \pm 0.01 M_\odot$ by fitting of
light curves.

The light curves are calculated for four companion masses, 
i.e., $M_{\rm MS}= 0.8$, 1.1, 1.5 and $2.0 M_\odot$.
Since we obtain similar light curves for all of these four cases,
we show here only the results for $M_{\rm MS}= 1.5 M_\odot$.
For a pair $1.37 M_\odot$ WD $+$ $1.5 M_\odot$ MS, for example,
we have the separation $a= 4.97 R_\odot$, 
the effective radius of the inner critical Roche lobe 
$R_1^*= 1.84 R_\odot$, and $R_2^*= R_2=1.92 R_\odot$, 
for the WD and MS (Fig. \ref{v394cra87_fig_burst}), respectively.

     In the plateau phase of the light curve ($t \sim 10$---30 days), 
i.e., when the WD photosphere is much smaller than the binary size,
the light curve is determined mainly by the irradiations of 
the ACDK and the MS as shown in the 1999 U Sco outburst (\cite{hkkm00}).
Here, we assume that the surfaces of the MS and of
the ACDK emit photons as a black-body at a local 
temperature of the surfaces heated by the WD photosphere. 
We assume further that a half of the absorbed energy 
is emitted from the surfaces of the MS and of the ACDK
(50\% efficiency, i.e., $\eta_{\rm ir, MS}=0.5$, 
and $\eta_{\rm ir, DK}=0.5$),
while the other is carried into interior of the accretion disk 
and eventually brought into the WD.  The unheated surface
temperatures are assumed to be $T_{\rm ph, MS}= 5000$ K 
for the MS and to be $T_{\rm ph, disk}= 4000$ K
for the ACDK including the disk rim.
The viscous heating of the ACDK is neglected during 
the outburst because it is much smaller than that of the irradiation 
effects.  We have checked different temperatures of
$T_{\rm ph, MS}= 4000$, 6000 K and $T_{\rm ph, disk}= 3000$, 5000 K,
but could not find any significant differences in the light curves.

\begin{figure}
\plotone{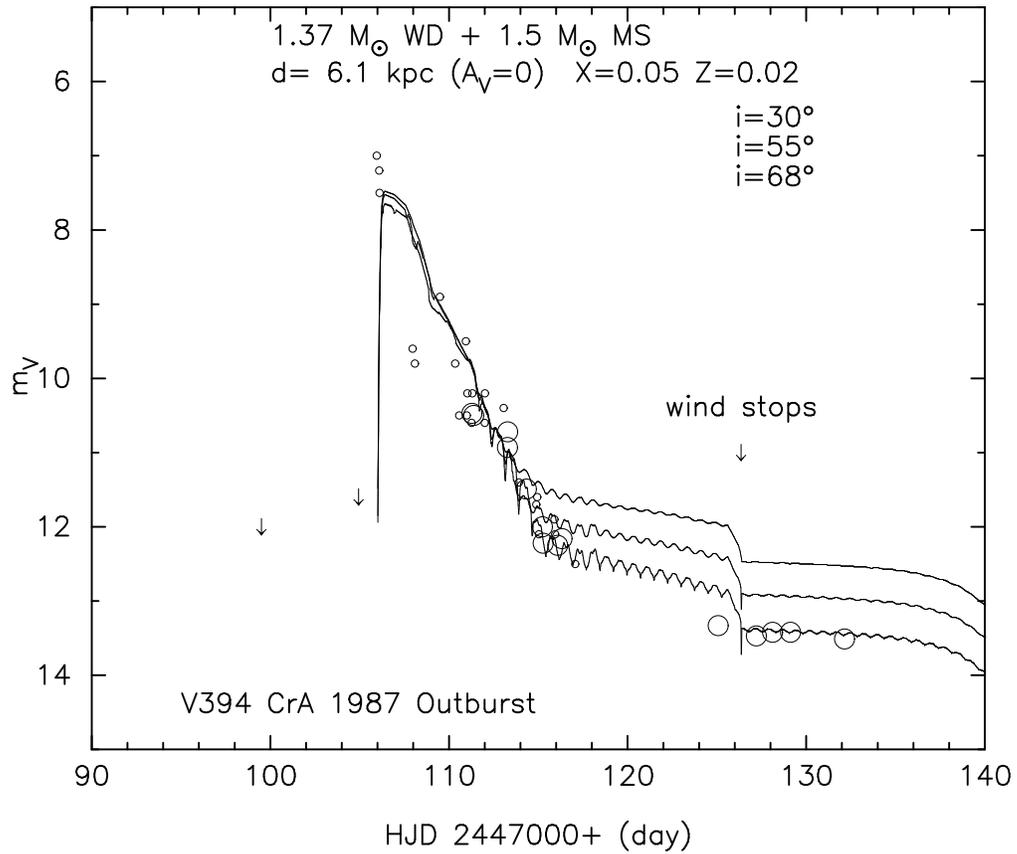}
\caption{
Theoretical $V$ light curves (solid lines) plotted against time
(HJD 2447000$+$) for 1.37 $M_\odot$ WD $+$ 1.5 $M_\odot$ MS with 
inclination angles of $i=30\arcdeg$, $i=55\arcdeg$, and $i=68\arcdeg$ 
from top to bottom, together with the observational points
(the large open circles represent data taken from Sekiguchi et al.
[1989], the small open circles from IAU Cir. No. 4428---4430, 4445).
The best fit one is obtained for the inclination angle of 
$i=65\arcdeg$---$70\arcdeg$.  The WD photosphere shrinks from 
$\sim 0.1 R_\odot$ to $0.005 R_\odot$ in a day when the optically 
thick wind stops on HJD 2,447,127.  As a result, the calculated 
visual light curve drops there.
The distance to V394 CrA is estimated to be 6.1 kpc by fitting 
for no absorption ($A_V=0$).
\label{vmag1370va1_v394cra1987}}
\end{figure}

     The luminosity of the accretion disk depends strongly on 
both the thickness $\beta$ and the size $\alpha$ in the plateau phase.
We have examined a total of 160 cases for the set 
of $(\alpha, \beta)$, which is the product of 16 cases 
of $\alpha=$ 0.5---2.0 by 0.1 step
and 10 cases of $\beta=$ 0.05---0.50 by 0.05 step.
Here, we have adopted the same set of ($\alpha$, $\beta$) as of 
the U Sco 1999 outburst (\cite{hkkm00}), i.e.,
$\alpha=$1.4 (1.2), and $\beta=$0.30 (0.35), 
for the outburst wind phase (for the static phase).
The hydrogen content $X$ cannot be determined from the light curve
fitting mainly because the observation of the 1987 outburst does not
cover the final decay phase toward quiescence.  So we have adopted
the same value of $X=0.05$ as in the U Sco 1999 outburst 
(\cite{hkkm00}).  Then, the optically thick wind stops 20 days
after the optical maximum 
as shown in Figure \ref{vmag1370va1_v394cra1987}.
Because the inclination angle of the orbit is not known yet, 
we have fitted the light curve by changing
the inclination angle, i.e., $i=30\arcdeg$, 
$45\arcdeg$, $50\arcdeg$, $55\arcdeg$, $60\arcdeg$, $65\arcdeg$, 
$66\arcdeg$, $67\arcdeg$, $68\arcdeg$, 
$69\arcdeg$, and $70\arcdeg$.  
Some of them are plotted in Figure \ref{vmag1370va1_v394cra1987}.
The visual magnitude of the 1987 outburst 
can be reproduced when the inclination angle is between 
$65\arcdeg$ and $70\arcdeg$ except for the first 
day of the outburst.  It is almost certain that the visual light 
during the first day exceeds the Eddington limit
because our optically thick wind solutions do not produce 
super Eddington luminosities.

     Based on the best fitted solutions, we have estimated 
the envelope mass at the optical maximum 
as $\Delta M= 5.8 \times 10^{-6} M_\odot$,
which indicates a mass accretion rate of 
$\dot M_{\rm acc}= 1.5 \times 10^{-7} M_\odot$ yr$^{-1}$
during the quiescent phase between the 1949 and 1987 outbursts 
if no WD matter has been dredged up.
About 77\% ($4.5 \times 10^{-6} M_\odot$) of the envelope mass 
has been blown off in the outburst wind while the residual 23\% 
($1.3 \times 10^{-6} M_\odot$) has been left and added 
to the helium layer of the WD.  Thus, the net mass increasing rate 
of the WD is $\dot M_{\rm He}= 0.34 \times 10^{-7} M_\odot$ yr$^{-1}$.

     The distance to V394 CrA is estimated to be 6.1 kpc for
no absorption ($A_V=0$).  We discuss the distance to V394 CrA 
in more detail, in the next section, to solve the discrepancy of
the distance estimations between in quiescence and in bursting phases.

\placefigure{vmag1370va1_v394cra1987}

\begin{figure}
\plotone{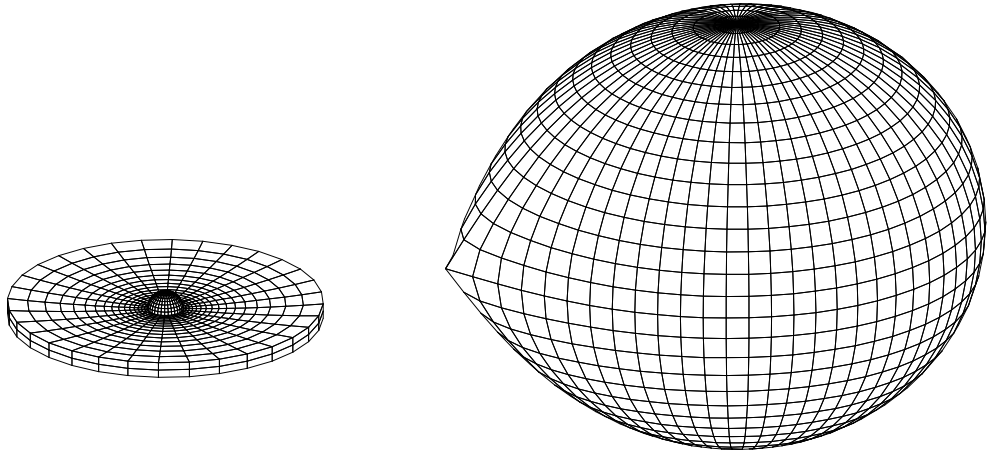}
\caption{
Configuration of our V394 CrA model at the quiescent phase.
The inclination angle is $i = 68\arcdeg$.  The disk size and
thickness are $\alpha=0.7$ and $\beta=0.05$, respectively.
The photospheric radius of the hot component 
is exaggerated in this figure to easily see it. 
\label{v394cra_fig_quiescent}}
\end{figure}

\placefigure{v394cra_fig_quiescent}

\section{LIGHT CURVES IN QUIESCENCE}
     In a quiescent phase, we have adopted the same binary model 
in the 1987 outburst phase except for the disk shape 
(see Fig. \ref{v394cra_fig_quiescent}), that is, we assume
the WD mass as $M_{\rm WD}= 1.37 M_\odot$, 
the accretion luminosity of the WD as
\begin{equation}
L_{\rm WD} = {1 \over 2} {{G M_{\rm WD} \dot M_{\rm acc}} 
\over {R_{\rm WD}}} + L_{\rm WD,0},
\label{accretion-luminosity}
\end{equation}
(e.g., \cite{sta88}) and the viscous luminosity and the irradiation 
effect of the ACDK as
\begin{equation}
\sigma T_{\rm ph, disk}^4 = {{3 G M_{\rm WD} \dot M_{\rm acc}} 
\over {8 \pi \varpi^3}} 
+ \eta_{\rm ir, DK} {{L_{\rm WD}} \over {4 \pi r^2}} \cos\theta,
\end{equation}
(e.g., \cite{sch97}), where 
$L_{\rm WD}$ and $L_{\rm WD,0}$ are the total and intrinsic 
luminosities of the WD, respectively,
$G$ the gravitational constant, $\dot M_{\rm acc}$ is
the mass accretion rate of the WD, 
$R_{\rm WD}= 0.0032 R_\odot$ is the radius of the $1.37 M_\odot$ WD,
$\sigma$ is the Stefan-Boltzmann constant,
$T_{\rm ph, disk}$ is the surface temperature of the ACDK, 
$r$ is the distance from the center of the WD, 
and $\cos\theta$ is the incident angle of the surface.
The accretion luminosity of the WD is as large as $\sim 1000 L_\odot$ 
for $\dot M_{\rm acc} \sim 1.5 \times 10^{-7} M_\odot$ yr$^{-1}$. 
The unheated temperatures are assumed to be 4000 K at the disk rim
and 5000 K at the MS photosphere.  

     Figure \ref{mix_lum_bv} shows the observational 
points (open circles) by Schaefer (1990) together with 
our calculated $B$ light curve 
(thick solid line) for the suggested mass accretion rate of
To fit our theoretical light curves with Schaefer's (1990) 
observational points, we have calculated $B$ light 
curves by changing the parameters of
$\alpha=0.5$---1.0 by 0.1 step, $\beta=0.05$---0.50 by 0.05 step,
$T_{\rm ph, MS}= 3000$---6000 K by 1000 K step,
$T_{\rm ph, disk}= 3000$ and 4000 K at the disk rim,
$L_{\rm WD,0}=0$---1000 $L_\odot$ by 100 $L_\odot$ step and
1000---5000 $L_\odot$ by 1000 $L_\odot$ step 
and $i=60$---$70\arcdeg$ by $1\arcdeg$ step and seek for
the best fit model for each mass accretion rate.
The best fit parameters obtained are shown in the figures
(also see Table \ref{v394cra_quiescence} for the other mass
accretion rates of (0.1---5.0)$\times 10^{-7} M_\odot$ yr$^{-1}$).

     Then we have calculated the theoretical color index $(B-V)_c$ 
for these best fit models.  Here, we explain only the case of
$\dot M_{\rm acc}= 1.5 \times 10^{-7} M_\odot$ yr$^{-1}$.
because the 1987 outburst model suggests 
an average mass accretion rate of 
$\sim 1.5 \times 10^{-7} M_\odot$ yr$^{-1}$
during the quiescent phase between the 1949 and 1987 outbursts. 
By fitting, we obtain the apparent distance 
modulus of $m_{B, 0}= 17.66$, which corresponds 
to the distance of 34 kpc without absorption ($A_B=0$).  
On the other hand, we obtained a rather blue color index 
of $(B-V)_c= -0.16$ at $m_B= 19.2$.  
This suggests a large color excess of 
$E(B-V)= (B-V)_o - (B-V)_c= 1.10$
with the observed color of $(B-V)_o= 0.94$ at
$m_B=19.20$ by Schaefer (1990).
Here, the suffixes $c$ and $o$ represent the theoretically 
calculated values and the observational values, respectively. 
Then, we have a large absorption of $A_V= 3.1 ~E(B-V)= 3.46$
and $A_B= A_V + E(B-V) = 4.56$.  
Thus, we obtain the distance of 4.2 kpc.
Then, V394 CrA lies $\sim 600$ pc below the Galactic plane
($l=352.84\arcdeg$, $b=-7.72\arcdeg$).

     The distance of 4.2 kpc indicates an absorption 
of $A_V= 0.81$ during the outburst (6.1 kpc for $A_V=0$), 
i.e., $A_V= 5 \log(6.1/4.2) = 0.81$.  
Then, we have $E(B-V)=A_V/3.1=0.26$ and $A_B= A_V + E(B-V)= 1.07$. 
On the other hand, Duerbeck (1988)
suggested an absorption of $A_B= 1$ from the nearby 
distance-interstellar absorption relation 
by Neckel and Klare (1980), which is consistent with
our estimation of the absorption during the 1987 outburst.
Thus, we may suggest that the intrinsic absorber of V394 CrA 
is blown off during the outburst as discussed in U Sco
(\cite{hkkmn00}).

\placetable{v394cra_quiescence}
\placefigure{mix_lum_bv}

\section{DISCUSSION}
     Even for much different mass accretion rates, the distance to 
V394 CrA has been estimated not to be so much different from 
4.2 kpc as tabulated in Table \ref{v394cra_quiescence}.
For lower mass accretion rates such as 
$\dot M_{\rm acc}= 1.0 \times 10^{-7} M_\odot$ yr$^{-1}$, however,
we need the 100\% irradiation efficiency of the MS 
($\eta_{\rm ir,MS}=1.0$) or an intrinsic luminosity of the WD 
as large as $L_{\rm WD,0} \sim 300 L_\odot$ 
for the 50\% efficiency ($\eta_{\rm ir,MS}=0.5$),
in order to reproduce the $\sim 0.5$ mag sinusoidal variation.
We also need an intrinsic luminosity of the WD as large as 
300 $L_\odot$ 
both for $\dot M_{\rm acc}= 0.5 \times 10^{-7} M_\odot$ yr$^{-1}$
and for $\dot M_{\rm acc}= 0.25 \times 10^{-7} M_\odot$ yr$^{-1}$,
and 200 $L_\odot$ 
for $\dot M_{\rm acc}= 0.1 \times 10^{-7} M_\odot$ yr$^{-1}$,
as summarized in Table \ref{v394cra_quiescence}.

     The brightness of the system depends on various model 
parameters adopted here, that is, the efficiency of 
the irradiations $\eta_{\rm ir, DK}$ and $\eta_{\rm ir, MS}$,
the intrinsic luminosity of the WD $L_{\rm WD,0}$, the power of
the disk shape $\nu$.  However, the distance estimation itself 
is hardly affected even if we introduce the different values of 
the parameters, as clearly shown in Table \ref{tbl-2}.
Thus, we may conclude that the determination of the distance 
to V394 CrA in quiescence is rather robust as has already been 
shown in U Sco (\cite{hkkm00}, 2000b). 

\begin{figure}
\plotone{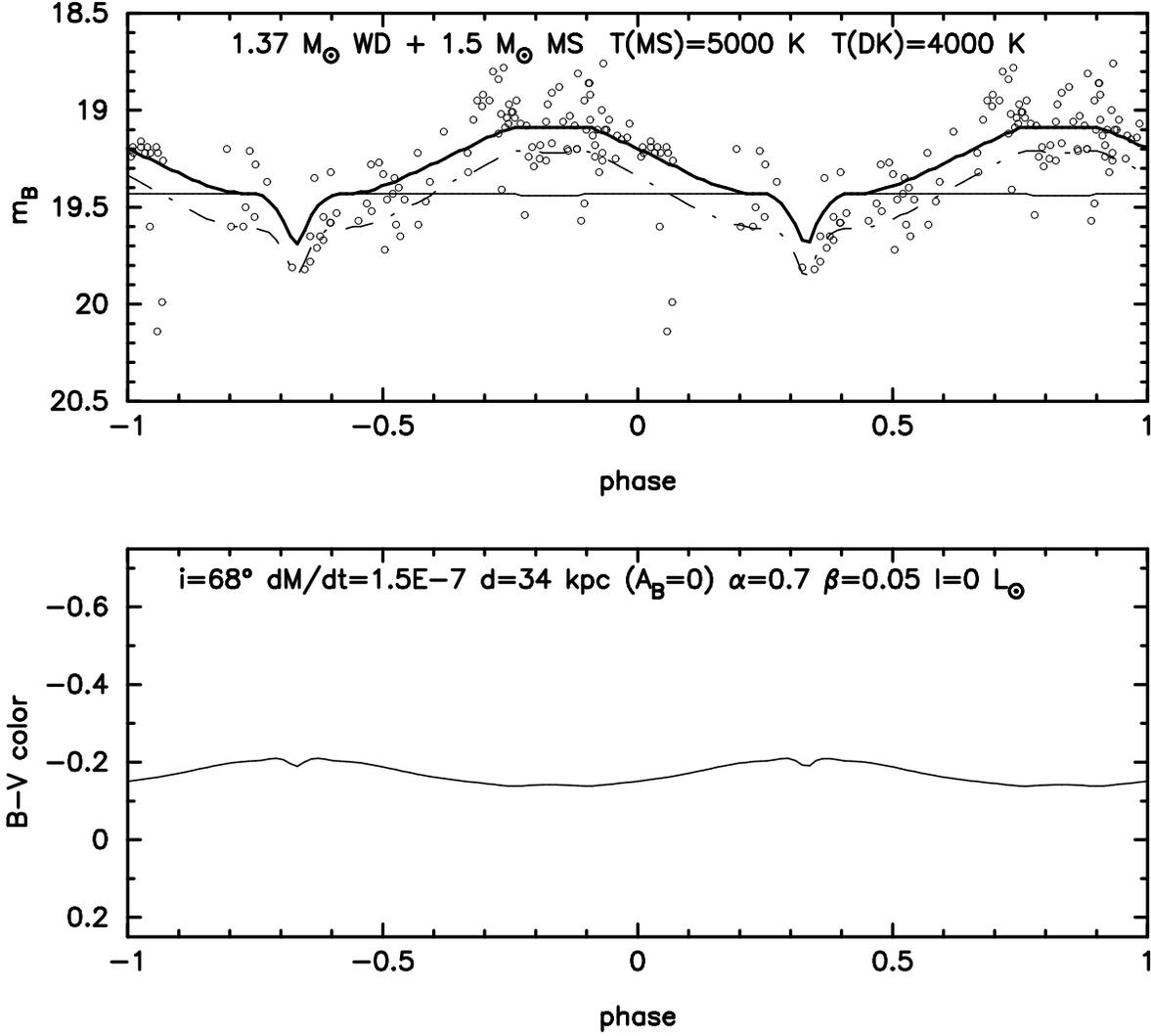}
\caption{
{\it Top}: theoretical $B$ light curves (thick solid lines)
plotted against the binary phase (two phases from $-1.0$ to $1.0$)
together with the observational points 
(open circles represent data taken from Schaefer 1990).
{\it Bottom}: theoretical $B-V$ color light curves (solid lines)
with no observational data points.
Two additional light curves (thin solid: non-irradiated MS, and
dot-dashed: non-irradiated ACDK) are also plotted to show
each contribution to the total $B$-light. 
We assume the mass accretion rate of
$1.5 \times 10^{-7}M_\odot$ yr$^{-1}$.
The other parameters are printed in the figures.  
About 0.5 mag sinusoidal variation can be reproduced with
the inclination angle between $i=$65$\arcdeg$---68$\arcdeg$ but
not with $i < 60\arcdeg$ because of too shallow variations. 
\label{mix_lum_bv}}
\end{figure}

     About 0.5 mag sinusoidal variation of the $B$
light curve during the quiescence needs a relatively large
reflection of the companion star as calculated 
in Figure \ref{mix_lum_bv}, thus indicating a relatively large
luminosity of the WD photosphere.  If the intrinsic luminosity
of the WD is negligibly small compared with the accretion 
luminosity (e.g., the nuclear burning is smaller than 
the accretion luminosity), the mass accretion rate should 
be higher than 
$\dot M_{\rm acc} \gtrsim 1.0 \times 10^{-7} M_\odot$ yr$^{-1}$
because the efficiency of the irradiation effect must be
smaller than 100\%, which is consistent with our estimation
of $\dot M_{\rm acc}= 1.5 \times 10^{-7} M_\odot$ yr$^{-1}$ 
derived from the envelope mass at the optical maximum.

     These systems with relatively high mass accretion rates are 
exactly the same as those proposed 
by Hachisu et al. (1999b) as a progenitor system of SNe Ia  
(see also \cite{lih97}).
Using the same simplified evolutional model as described 
in Hachisu et al. (1999b), we have followed binary evolutions 
for various pairs with the initial sets of 
($M_{1,i}$, $M_{2,i}$, $a_i$), 
i.e., for the initial primary masses of $M_{1,i}=4$, 5, 6, 7, 
and $9 M_\odot$, the initial secondary masses of 
$M_{2,i}=1.7$---$3.0 M_\odot$ by $\Delta M_{2,i}=0.1 M_\odot$ step,
and the initial separations of $a_i=80$---$600 R_\odot$ 
by $\Delta \log a_i= 0.01$ step.  
Starting from the initial set
($7 M_\odot$, $2.0 M_\odot$, $150 R_\odot$),
for example, we have obtained a binary system of
$M_{\rm WD,0}= 0.9 M_\odot$, $M_{\rm MS,0}= 2.2 M_\odot$,
and $P_0= 1.375$ days, after the binary underwent the first common 
envelope evolution and then the primary naked helium star evolved 
to a helium giant and had transferred helium to the secondary MS.

\placetable{tbl-2}

     Then, the secondary MS has slightly evolved to expand and
filled its Roche lobe.  Mass transfer begins from the MS to the WD.  
We have further followed evolution of the binary
until the binary reaches $M_{\rm WD}= 1.37 M_\odot$ 
and $P= 0.7577$ days at the same time, that is,
we regard the binary as V394 CrA
when both the conditions, $M_{\rm WD}= 1.37 M_\odot$ 
and $P= 0.7577$ days, are satisfied at the same time.  
Then, we obtain the present state of V394 CrA having the secondary mass
of $M_{\rm MS}= 1.39 M_\odot$ and the mass transfer rate of 
$\dot M_2= 1.6 \times 10^{-7} M_\odot$ yr$^{-1}$. 
In our evolutionary model, this binary system will soon explode 
as an SN Ia when the WD mass reaches $M_{\rm Ia}= 1.378 M_\odot$. 
The mass transfer rate of our evolutionary model is consistent with 
$\sim 1.5 \times 10^{-7} M_\odot$ yr$^{-1}$ 
estimated from the light curve fitting.

     Finally, we may conclude that V394 CrA is the second
strong candidate for Type Ia progenitors, next to U Sco 
(Hachisu et al. 2000a, 2000b).

\acknowledgments
This research has been supported in part by the Grant-in-Aid for
Scientific Research (09640325, 11640226) 
of the Japanese Ministry of Education, Science, Culture, and Sports.

%
%

\clearpage

\clearpage
\begin{deluxetable}{cccccccccc}
\footnotesize
\tablecaption{V394 CrA quiescent phase\tablenotemark{a}.
\label{v394cra_quiescence}}
\tablewidth{0pt}
\tablehead{
\colhead{$\dot M_{\rm acc}$} &
\colhead{$L_{\rm WD,0}$} &
\colhead{$\beta$} &
\colhead{$\eta_{\rm ir,MS}$} &
\colhead{$m_{B,0}$} &
\colhead{$(B-V)_c$} &
\colhead{$E(B-V)$} &
\colhead{$A_V$} &
\colhead{$A_B$} &
\colhead{$d$} \nl
\colhead{($M_\odot$ yr$^{-1}$)} &
\colhead{($L_\odot$)} &
\colhead{} &
\colhead{} &
\colhead{} &
\colhead{at $B=19.2$} &
\colhead{at $B=19.2$} &
\colhead{} &
\colhead{} &
\colhead{(kpc)}
} 
\startdata
5.0$\times 10^{-7}$ & 0.0 & 0.10 & 0.5 & 18.45 & $-0.28$ 
& 1.22 & 3.83 & 5.05 & 4.8 \nl
2.5$\times 10^{-7}$ & 0.0 & 0.05 & 0.5 & 17.96 & $-0.20$
& 1.14 & 3.58 & 4.72 & 4.4 \nl
1.5$\times 10^{-7}$ & 0.0 & 0.05 & 0.5 & 17.66 & $-0.16$
& 1.10 & 3.46 & 4.56 & 4.2 \nl
1.0$\times 10^{-7}$ & 0.0 & 0.05 & 1.0 & 17.49 & $-0.11$
& 1.05 & 3.30 & 4.35 & 4.3 \nl
1.0$\times 10^{-7}$ & 300 & 0.05 & 0.5 & 17.53 & $-0.12$
& 1.06 & 3.33 & 4.39 & 4.2 \nl
5.0$\times 10^{-8}$ & 300 & 0.05 & 0.5 & 17.16 & $-0.06$
& 1.00 & 3.14 & 4.14 & 4.0 \nl
2.5$\times 10^{-8}$ & 300 & 0.05 & 0.5 & 16.86 & $+0.00$
& 0.94 & 2.95 & 3.89 & 3.9 \nl
1.0$\times 10^{-8}$ & 200 & 0.05 & 0.5 & 16.39 & $+0.11$
& 0.83 & 2.61 & 3.44 & 3.9 \nl
\enddata
\tablenotetext{a}{Here, the inclination angle of the orbit 
$i=68\arcdeg$, the disk size $\alpha=0.70$,
the efficiency of the ACDK irradiation $\eta_{\rm ir,DK}=0.5$,
the original temperatures of the companion $T_{\rm ph, MS}=5000$ K and
of the disk rim $T_{\rm ph, disk}= 4000$ K are assumed for all cases.
}
\end{deluxetable}

\begin{deluxetable}{lcccccccc}
\footnotesize
\tablecaption{Model dependence of the distance.
\label{tbl-2}}
\tablewidth{0pt}
\tablehead{
\colhead{model\tablenotemark{a}} &
\colhead{$L_{\rm WD,0}$} &
\colhead{$\beta$} &
\colhead{$m_{B,0}$} &
\colhead{$(B-V)_c$}&
\colhead{$E(B-V)$} &
\colhead{$A_V$} &
\colhead{$A_B$} &
\colhead{$d$} \nl
\colhead{} &
\colhead{$(L_\odot)$} &
\colhead{} &
\colhead{} &
\colhead{at $B=19.2$} &
\colhead{at $B=19.2$} &
\colhead{} &
\colhead{} &
\colhead{(kpc)}
} 
\startdata
$L_{\rm WD,0}=1000 L_\odot$ & 1000 & 0.10 & 18.06 & $-0.21$ 
& 1.15 & 3.61 & 4.76 & 4.6 \nl
$L_{\rm WD,0}=2000 L_\odot$ & 2000 & 0.12 & 18.27 & $-0.24$ 
& 1.18 & 3.71 & 4.88 & 4.7 \nl
$\eta_{\rm ir,MS}=1.0$ & 0 & 0.10 & 17.84 & $-0.17$ 
& 1.11 & 3.49 & 4.60 & 4.5 \nl
$\eta_{\rm ir,MS}=0.25$ & 2000 & 0.05 & 18.01 & $-0.19$ 
& 1.13 & 3.55 & 4.68 & 4.6 \nl
$\eta_{\rm ir,DK}=1.0$ & 500 & 0.05 & 18.06 & $-0.19$ 
& 1.13 & 3.55 & 4.68 & 4.7 \nl
$\eta_{\rm ir,DK}=0.25$ & 300 & 0.05 & 17.66 & $-0.15$ 
& 1.09 & 3.42 & 4.51 & 4.3 \nl
$\nu=3.0$ & 0 & 0.05 & 17.69 & $-0.16$ 
& 1.10 & 3.46 & 4.56 & 4.2 \nl
$\nu=1.25$ & 0 & 0.05 & 17.53 & $-0.14$ 
& 1.08 & 3.39 & 4.47 & 4.1 \nl
\enddata
\tablenotetext{a}{inclination angle $i=68\arcdeg$, 
mass accretion rate $\dot M_{\rm acc}= 1.5 \times 10^{-7} 
M_\odot$ yr$^{-1}$, disk parameters $\alpha=0.7$ and $\beta=0.05$,
disk surface parameter $\nu=2$, irradiation efficiencies 
$\eta_{\rm ir, MS}=0.5$ and $\eta_{\rm ir, DK}=0.5$,
intrinsic luminosity of the WD $L_{\rm WD,0}=0$
 for all cases, otherwise specified.
}
\end{deluxetable}

%
%
%
%


\clearpage

\clearpage

\clearpage

\clearpage

%

\end{document}